\documentclass[aps,prd,reprint,twocolumn,superscriptaddress,preprintnumbers,nofootinbib]{revtex4-1}

\usepackage{amsthm}
\usepackage{amsmath}
\usepackage{graphicx}
\usepackage{slashed}
\usepackage{amssymb}
\usepackage{float}
\usepackage[utf8]{inputenc}
\usepackage[T1]{fontenc}
\usepackage[colorlinks=True, citecolor=blue, urlcolor=blue, linkcolor=blue]{hyperref}
\usepackage{xcolor}
\usepackage{bbm}
\usepackage{bbold}

\newcommand*\diff{\mathop{}\!\mathrm{d}}

\newcommand{\nn}{\nonumber}

\newcommand{\be}{\begin{eqnarray}}
\newcommand{\ee}{\end{eqnarray}}
\newcommand{\ma}{\mathrm}
\newcommand{\ml}{\mathcal}
\newcommand{\bs}{\boldsymbol}
\newcommand{\Tr}{\mathrm{Tr}}

\begin{document}

\title{Real Time Quarkonium Transport Coefficients in Open Quantum Systems from Euclidean QCD}

\author{Bruno Scheihing-Hitschfeld}
\email{bscheihi@mit.edu}
\affiliation{Center for Theoretical Physics, Massachusetts Institute of Technology, Cambridge, MA 02139, USA}

\author{Xiaojun Yao}
\email{xjyao@uw.edu}
\affiliation{InQubator for Quantum Simulation, Department of Physics, University of Washington, Seattle, WA, 98195, USA}

\date{\today}
\preprint{INT-PUB-23-019, MIT-CTP/5575, IQuS@UW-21-051}
\begin{abstract}
Recent open quantum system studies showed that quarkonium time evolution inside the quark-gluon plasma is determined by transport coefficients that are defined in terms of a gauge invariant correlator of two chromoelectric field operators connected by an adjoint Wilson line. We study the Euclidean version of the correlator for quarkonium evolution and discuss the extraction of the transport coefficients from this Euclidean correlator, highlighting its difference from other problems that also require reconstructing a spectral function, such as the calculation of the heavy quark diffusion coefficient. Along the way, we explain why the transport coefficient $\gamma_{\rm adj}$ differs from $\gamma_{\rm fund}$ at finite temperature at $\mathcal{O}(g^4)$, in spite of the fact that their corresponding spectral functions differ only by a temperature-independent term at the same order. We then discuss how to evaluate the Euclidean correlator via lattice QCD methods, with a focus on reducing the uncertainty caused by infrared renormalons in determining the renormalization factor nonperturbatively.
\end{abstract}

\maketitle

\section{Introduction}
The scientific mission of relativistic heavy ion colliders is to investigate properties of the deconfined phase of nuclear matter in the high temperature regime, known as the quark-gluon plasma (QGP). In current heavy ion collision experiments, the QGP only lives for a short time period (roughly $10$ fm/c in the laboratory frame) and we cannot directly measure its properties. Therefore, we use probes such as particle multiplicities and azimuthal distributions, jets and hadrons containing heavy quarks to indirectly study its properties. Various properties of the QGP are encoded in terms of gauge invariant correlation functions of field operators that often define transport coefficients showing up in the time evolution equations of the probes in the medium. Well-known examples include the shear viscosity (defined as a correlator of stress-energy tensors), the jet quenching parameter (a correlator of light-like Wilson lines) and the heavy quark diffusion coefficient (a correlator of two chromoelectric fields dressed with Wilson lines). Since the QGP is a strongly coupled fluid, nonperturbative determinations of these transport coefficients are crucial in our understanding of the QGP and QCD at finite temperature. Common nonperturbative methods include lattice QCD calculations and the holographic correspondence~\cite{Casalderrey-Solana:2011dxg}. One can also extract these transport coefficients from experimental data by solving in-medium evolution equations (which can be model dependent) for different values of the transport coefficients and then performing a Bayesian analysis~\cite{Bernhard:2019bmu,Nijs:2020ors,Nijs:2020roc,JETSCAPE:2020mzn,JETSCAPE:2021ehl}.

Recently, thanks to the advance in applying the open quantum system framework to study jets~\cite{Vaidya:2020cyi} and quarkonia~\cite{Akamatsu:2011se,Akamatsu:2014qsa,Katz:2015qja,Brambilla:2016wgg,Brambilla:2017zei,Blaizot:2017ypk,Kajimoto:2017rel,Blaizot:2018oev,Yao:2018nmy,Akamatsu:2018xim,Miura:2019ssi,Sharma:2019xum,Yao:2020eqy,Akamatsu:2021vsh,Brambilla:2021wkt,Miura:2022arv,Brambilla:2022ynh,Xie:2022tzs,alalawi2023impact} in the QGP (for recent reviews, see Refs.~\cite{Rothkopf:2019ipj,Akamatsu:2020ypb,Sharma:2021vvu,Yao:2021lus}), a novel correlator of two chromoelectric fields dressed with Wilson lines that determines transport properties of quarkonium in the medium was constructed~\cite{Brambilla:2017zei,Yao:2020eqy}. This correlator for quarkonium transport is similar to but different from the correlator defining the heavy quark diffusion coefficient~\cite{Casalderrey-Solana:2006fio,Caron-Huot:2009ncn} in terms of the ordering of the fields contained in the Wilson lines. Perturbative calculations in $R_\xi$ gauge showed that the spectral function of the correlator for quarkonium transport~\cite{Binder:2021otw} differs from that for heavy quark transport~\cite{Burnier:2010rp} by a temperature independent constant at next-to-leading order (NLO). However, if both calculations had been performed in temporal axial gauge ($A_0=0$), one would, at first sight, have concluded that the two correlators were identical. This resulted in a puzzle: Since both correlators are defined in a gauge invariant way, calculations with different gauge choices must give the same result. This puzzle was resolved in Ref.~\cite{Scheihing-Hitschfeld:2022xqx}, establishing the difference between the two correlators on a more solid ground in QCD. Beyond NLO, the heavy quark diffusion coefficient has been studied by using hard-thermal-loop resummation~\cite{Caron-Huot:2007rwy}, as well as nonperturbatively via the lattice QCD method~\cite{Banerjee:2011ra,Ding:2012sp,Francis:2015daa,Brambilla:2020siz,Altenkort:2020fgs} and the AdS/CFT correspondence~\cite{Herzog:2006gh,Gubser:2006qh,Casalderrey-Solana:2006fio}. On the other hand, a recent AdS/CFT calculation showed that the analog quarkonium transport coefficients in $\mathcal{N}=4$ supersymmetric Yang-Mills (SYM) theory are zero~\cite{Nijs:2023dks}, in stark contrast to the heavy quark diffusion coefficient value of $ \sqrt{\lambda} \pi T^3$ at large coupling $\lambda = g^2 N_c \gg 1$. This difference is surprising because the heavy quark and quarkonium transport coefficients are defined by similar chromoelectric field correlators. Therefore, it is well motivated to study the quarkonium transport properties nonperturbatively in QCD. It is also crucial and urgent, since quarkonium production serves as an important probe of the QGP that is produced strongly coupled in current heavy ion collision experiments.

In this article, we discuss how to extract the quarkonium transport coefficients from lattice QCD calculations of a specific Euclidean chromoelectric correlator. The paper is organized as follows: We will first review the quarkonium transport coefficients in the real-time formalism in Section~\ref{sect:EE}, which are defined in terms of a correlator of two chromoelectric fields connected via an adjoint Wilson line. Then, in Section~\ref{sect:euclidean} we will discuss the Euclidean version of the correlator and how to relate it to its real time counterpart. Next, in Section~\ref{sect:lattice} the setup of a lattice QCD calculation of this Euclidean correlator will be discussed, with a focus on how to renormalize it. Finally, we will conclude and present our outlook in Section~\ref{sect:conclusion}.

\section{Quarkonium Transport Properties}
\label{sect:EE}
The quarkonium transport coefficients are defined in terms of time-ordered chromoelectric field operators, dressed with Wilson lines~\cite{Brambilla:2017zei}:
\begin{align}
\label{eqn:adj}
\kappa_{\rm adj} &\equiv \frac{ g^2 T_F}{3 N_c} {\rm Re} \int \diff t\, \big\langle \ml{T} E^a_i(t) W^{ab}(t,0) E^b_i(0) \big\rangle_T \\
\gamma_{\rm adj} &\equiv \frac{ g^2 T_F}{3 N_c} {\rm Im} \int \diff t\, \big\langle \ml{T} E^a_i(t) W^{ab}(t,0) E^b_i(0) \big\rangle_T \,,\nn
\end{align}
where $\langle O \rangle_T \equiv \Tr(O\,e^{-\beta H})/\Tr(e^{-\beta H})$, $E^a_i$ is a chromoelectric field, $W^{ab}(t,0)$ denotes a time-like Wilson line in the adjoint representation from $t=0$ to $t$, $\ml{T}$ represents the time-ordering symbol, $N_c = 3$ is the number of colors, and $T_F = 1/2$ is the normalization of the fundamental representation generator matrices. To simplify the notation we have neglected the spatial coordinates, which are the same for all the fields, and will do so throughout the paper, unless the spatial coordinates are no longer the same. Both $\kappa_{\rm adj}$ and $\gamma_{\rm adj}$ appear in the Lindblad equation describing the time evolution of a heavy quark-antiquark pair ($Q\bar{Q}$) at a small distance in the quantum Brownian motion limit~\cite{Brambilla:2016wgg,Brambilla:2017zei}:
\begin{align}
\label{eqn:lindblad}
\frac{\diff \rho_S(t)}{\diff t} = & -i\big[ H_S + \gamma_{\rm adj} \Delta h_S,\, \rho_S(t) \big] \\& +  \kappa_{\rm adj} \big( L_{\alpha i} \rho_S(t) L^\dagger_{\alpha i} - \frac{1}{2}\{  L^\dagger_{\alpha i}L_{\alpha i},\, \rho_S(t)\} \big)\,, \nn
\end{align}
where $\rho_S$ is the subsystem density matrix of the $Q\bar{Q}$ pair, $\gamma_{\rm adj} \Delta h_S$ is the thermal correction to the vacuum $Q\bar{Q}$ Hamiltonian $H_S$ and $L_{\alpha i}$ denotes the relevant Lindblad ``jump'' operators. Their explicit expressions are given in Appendix~\ref{app:lindblad}. The $\kappa_{\rm adj}$ parameter in the non-Hermitian part of the Lindblad equation determines the rate of transition between a $Q\bar{Q}$ pair in the color singlet state and that in the color octet state, as well as wavefunction decoherence. On the other hand, the $\gamma_{\rm adj}$ parameter in the Hermitian part of the Lindblad equation controls the modification of the $Q\bar{Q}$ potential caused by the medium.

One way to interpret the integrations in Eq.~\eqref{eqn:adj} is as Fourier transforms that convert the time domain to the frequency domain. Consequently, the coefficients $\kappa_{\rm adj}$ and $\gamma_{\rm adj}$ are the zero frequency limits of frequency-dependent correlation functions. Moreover, their behavior at finite frequency also turns out to be physically important. To explain the physical meaning of these correlation functions at finite frequency, we introduce path-ordered chromoelectric field correlators~\cite{Yao:2020eqy,Binder:2021otw}
\begin{align} 
\label{eq:g-definitions}
[g_{\rm adj}^{++}]^>(t) &\equiv \frac{g^2 T_F }{3 N_c}  \big\langle E_i^a(t)W^{ac}(t,+\infty) \\ & \qquad \qquad  \qquad \qquad \quad
W^{cb}(+\infty,0) E_i^b(0) \big\rangle_T \nn\\
[g_{\rm adj}^{--}]^>(t) &\equiv \frac{g^2 T_F }{3 N_c} \big\langle W^{dc}(-i\beta - \infty, -\infty)
W^{cb}(-\infty,t) \nn \\ & \quad \quad \quad \quad \quad \quad  E_i^b(t)
E_i^a(0)W^{ad}(0,-\infty)  \big\rangle_T  \, , \nn
\end{align}
and consider their Fourier transforms $[g_{\rm adj}^{\pm\pm}]^>(\omega) = \int \diff t\, e^{i\omega t} [g_{\rm adj}^{\pm\pm}]^>(t)$. The path-ordered version is more convenient to use at finite frequency and is consistent with the time-ordered version: It has been shown that $\kappa_{\rm adj} = [g_{\rm adj}^{++}]^>(\omega=0)$~\cite{Scheihing-Hitschfeld:2022xqx}. We note that because of the explicit operator ordering, only in $[g_{\rm adj}^{++}]^>$ the adjoint Wilson lines can be rewritten as $W^{ab}(t,0)$, which appears in the time-ordered correlator shown in Eq.~\eqref{eqn:adj}. Furthermore, one can obtain the time-ordered correlator that enters the definition of $\kappa_{\rm adj}$ and $\gamma_{\rm adj}$ by considering 
\begin{align}
\label{eqn:g>_to_gT}
[g_{\rm adj}^{{\mathcal{T}}}](t) &\equiv \big\langle \ml{T} E^a_i(t) W^{ab}(t,0) E^b_i(0) \big\rangle_T \nn\\
&= \theta(t) [g_{\rm adj}^{++}]^>(t) + \theta(-t) [g_{\rm adj}^{++}]^>(-t) \, .
\end{align}
The path-ordered correlators at finite frequency appear in the Boltzmann (rate) equation for quarkonium dissociation and recombination, which is derived in the quantum optical limit of the open quantum system approach~\cite{Yao:2020eqy,Binder:2021otw}:
\begin{align}
\label{eqn:rate}
\frac{\diff n_b(t,{\bs x})}{\diff t} = -\Gamma\, n_b(t,{\bs x}) + F(t,{\bs x}) \,,
\end{align}
where $n_b(t,{\bs x})$ is the density of the quarkonium state $b$ at time $t$, $\Gamma$ denotes the dissociation rate and $F$ represents the formation of the quarkonium state $b$ from a recombining pair of unbound heavy quarks $Q\bar{Q}$
\begin{align}
\label{eqn:disso}
\Gamma &=  \int \frac{\diff^3p_{\ma{rel}}}{(2\pi)^3} 
| \langle \psi_b | {\bs r} | \Psi_{{\bs p}_\ma{rel}} \rangle |^2 [g^{++}_E]^{>}(-\Delta E) \\
\label{eqn:reco}
F &=  \int \frac{\diff^3p_{\ma{cm}}}{(2\pi)^3}  \frac{\diff^3p_{\ma{rel}}}{(2\pi)^3} 
| \langle \psi_b | {\bs r} | \Psi_{{\bs p}_\ma{rel}} \rangle |^2 \\
& \quad \times
[g^{--}_E]^{>}(\Delta E) f_{Q\bar{Q}}(t, {\bs x}, {\bs p}_{\ma{cm}}, {\bs x}_{\rm rel}=0, {\bs p}_{\ma{rel}}) \,,  \nn
\end{align}
where a nonzero energy difference between the bound and unbound states $\Delta E=p^2_\ma{rel}/M+|E_{b}|$ determines how the finite frequency dependence of the correlators appears in the transition rates. Here $M$ is the heavy quark mass and $E_b$ is the binding energy of the quarkonium state $b$. The transition occurs via a color dipole interaction $\langle \psi_b | {\bs r} | \Psi_{{\bs p}_\ma{rel}} \rangle$ between a bound $Q\bar{Q}$ state $|\psi_b\rangle$ and an unbound scattering wave $|\Psi_{{\bs p}_\ma{rel}}\rangle$. $f_{Q\bar{Q}}$ denotes the distribution of unbound heavy quark pairs with center-of-mass positions $\bs x$ and momenta ${\bs p}_{\rm cm}$ and relative positions ${\bs x}_{\rm rel}=0$ and momenta ${\bs p}_{\rm rel}$.

The chromoelectric field correlator for quarkonium transport is different from that for heavy quark diffusion. In particular, the heavy quark diffusion coefficient $\kappa_{\rm fund}$ and an analogous quantity $\gamma_{\rm fund}$ (whose physical meaning has not been explored for heavy quark transport) are defined by
\begin{align}
\label{eqn:fund}
\kappa_{\rm fund} = \frac{g^2}{3N_c} {\rm Re} \! \int \! \diff t & \big\langle \Tr_{\rm c}[ U(-\infty,t)  \nn \\ & \qquad E_i(t) U(t,0) E_i(0) U(0, -\infty)  ] \big\rangle_{T,Q}  \\
\gamma_{\rm fund} = \frac{g^2}{3N_c} {\rm Im} \! \int \! \diff t &
\big\langle \Tr_{\rm c}[ U(-\infty,t)  \nn\\ & \qquad E_i(t) U(t,0) E_i(0) U(0, -\infty)] \big\rangle_{T,Q} \,, \nn
\end{align}
where $E_i = E_i^a T^a_F$ is the Lie algebra-valued chromoelectric field, with the fundamental representation generator matrices normalized as ${\rm Tr}_c(T^a_F T^b_F) = T_F \delta^{ab}$. Also, ${\rm Tr}_c$ denotes trace over color indices and $U(t,0)$ represents a time-like fundamental Wilson line from $t=0$ to $t$. The subscript $T$ in the expectation value denotes that the state on which this expectation value is calculated is a thermal density matrix, while the subscript $Q$ means that this thermal state contains a static external color charge in the fundamental representation, e.g., a heavy quark. Mathematically, this expectation value is defined as $\langle O\rangle_{T,Q} \equiv N_c {\rm Tr}[U(-i\beta - \infty, -\infty) O e^{-\beta H}]/{\rm Tr}[U(-i\beta - \infty, -\infty) e^{-\beta H}]$ and thus is different from that in Eq.~\eqref{eqn:adj}.
The fundamental Wilson line along the imaginary time at $t=-\infty$ indicates the inclusion of the heavy quark effect on the thermal density matrix of the whole system.
It is noted that the operators involved in the definition of $\kappa_{\rm fund}$ and $\gamma_{\rm fund}$ are path-ordered. We want to emphasize that the crucial difference between Eqs.~\eqref{eqn:adj} and~\eqref{eqn:fund} is not the representations of the Wilson lines, but the different orderings of the operators.

\section{Euclidean Correlators and Transport Coefficients}
\label{sect:euclidean}
As is well known, lattice QCD methods can only calculate correlation functions in Euclidean space and thus cannot be applied directly to study the real-time correlators defined in Eq.~\eqref{eq:g-definitions}.
In this section, we will introduce a Euclidean version of the correlator for quarkonium transport and discuss how to extract the quarkonium transport coefficients from the evaluation of this Euclidean correlator. As we will show, both the Euclidean correlator itself and the method to extract the quarkonium transport coefficients are different from the case of heavy quark diffusion in subtle and important aspects. To make the comparison more explicit, and also to take advantage of the apparent similarities between them, we will first review the extraction of the heavy quark diffusion coefficient from the corresponding Euclidean correlator. 

\subsection{Heavy Quark Diffusion}
The Euclidean correlator relevant for the heavy quark diffusion case is given by~\cite{Caron-Huot:2009ncn}
\begin{align}
\label{eqn:Gfund}
G_{\rm fund}(\tau) = - \frac{1}{3} \frac{\big\langle {\rm Re} {\rm Tr}_c[ U(\beta,\tau) gE_i(\tau) U(\tau,0) gE_i(0) ] \big\rangle_T}{\big\langle {\rm Re} {\rm Tr}_c[U(\beta,0)] \big\rangle_T} \,,
\end{align}
where $\beta=1/T$ is the inverse of the QGP temperature and $\langle\cdot\rangle_T = \Tr(\cdot e^{-\beta H})/\Tr(e^{-\beta H})$, with $H$ the Hamiltonian of the QGP in the absence of any external color source. It has been shown that the heavy quark transport coefficient can be obtained from $G_{\rm fund}(\tau)$ via~\cite{Caron-Huot:2009ncn,Eller:2019spw}
\begin{align}
\label{eqn:fund_from_rho}
\kappa_{\rm fund} &= \lim_{\omega\to0} \frac{T}{\omega} \rho_{\rm fund}(\omega) \, , \\
\gamma_{\rm fund} &= -\int_0^\beta \diff\tau \, G_{\rm fund}(\tau) \, ,\nn
\end{align}
where the spectral function $\rho_{\rm fund}(\omega)$ is related to the Euclidean correlator through\footnote{Our convention for the Fourier transform is $O(\omega) = \int \diff t e^{i\omega t}O(t)$.}
\begin{align}
\label{eqn:Gfund_rho}
G_{\rm fund}(\tau) = \int_0^{+\infty} \frac{\diff\omega}{2\pi}  \frac{\cosh\big(\omega(\tau-\frac{1}{2T})\big)}{\sinh\big( \frac{\omega}{2T} \big)} \rho_{\rm fund}(\omega) \,.
\end{align}
This correlator is constructed such that the standard Kubo-Martin-Schwinger (KMS) and analytic continuation relations hold as in textbook thermal field theory. Given an analytic expression for $\tilde{G}_{\rm fund}(\omega_n)$, with $\omega_n = 2\pi T n$, $n \in \mathbb{Z}$ the Matsubara frequencies, one can extract the spectral function by taking the real part\footnote{Many studies define correlation functions with an imaginary unit prefactor, and there the spectral function corresponds to the imaginary part of the retarded correlator, which has a factor of $1/2$ compared with the spectral function defined by the difference between the $>$ and $<$ Wightman correlators in frequency space.} of the retarded correlator obtained by analytic continuation $\omega_n \to -i (\omega + i\epsilon)$ of this Euclidean correlator. This has been done both at weak~\cite{Burnier:2010rp} (QCD) and strong~\cite{Casalderrey-Solana:2006fio} ($\mathcal{N}=4$ SYM) coupling. However, at physical values of the coupling in QCD, the only tool available at the moment is lattice gauge theory, and as such, the reconstruction of the spectral function $\rho_{\rm fund}$ through the relation~\eqref{eqn:Gfund_rho} has received much attention in recent years~\cite{Altenkort:2020fgs,Altenkort:2023oms,Brambilla:2022xbd}.

Comparatively, the theoretical treatment of quarkonium transport coefficients has received less attention. We now aim to fill in this gap, and subsequently, to provide a recipe to determine these transport coefficients from lattice QCD calculations. To this end, we need to first construct a Euclidean version of the correlator for quarkonium transport that can be calculated via lattice QCD methods, and then explain how to extract the quarkonium transport coefficients from the evaluation of such an Euclidean correlator. We will answer these two questions in the following two subsections. Details of the lattice calculation of the Euclidean correlator will be discussed in the next section.

\subsection{Euclidean Correlator for Quarkonium Transport} 
To construct the Euclidean correlator for quarkonium transport, we first note that because of the operator ordering in the definitions~\eqref{eq:g-definitions}, we can equivalently write
\begin{equation}
    [g_{\rm adj}^{++}]^>(t) = \frac{g^2 T_F }{3 N_c} \big\langle E_i^a(t)W^{ab}(t,0) E_i^b(0) \big\rangle_T \, .
\end{equation}
To perform the analytic continuation, it is best to explicitly isolate the $t$ dependence from the field operators and write it purely in terms of time evolution factors. We let $H$ be the Hamiltonian of the thermal bath QGP in the absence of any external color charge. When an external adjoint color charge is present, the Hamiltonian of the thermal bath is given by $[H \mathbbm{1} - g A_0^c(0) T^c_{\rm adj} ]^{ab}$.
The reason for the appearance of this modified Hamiltonian can be seen from converting the adjoint Wilson line back to the Schr\"odinger picture from the interaction picture
\be
e^{-i H t} W^{ab}(t,0) =  \left[e^{- i (H - g A_0^c(0) T_{\rm adj}^c ) t}\right]^{ab} \,. \label{eq:W-as-eiHt}
\ee
Eq.~\eqref{eq:W-as-eiHt} has the following physical interpretation: during the time interval between $0$ and $t$ the QGP evolves in the presence of an adjoint color charge, which is manifest in the modification of the Hamiltonian by $- g A_0$. It is essentially a local modification to Gauss's law\footnote{An interesting question one can ask of this expression is whether we still have explicit gauge invariance. The answer is, naturally, affirmative. However, this is not as easy to see when considering time-dependent gauge transformations as it is for time-independent gauge transformations. This is because the Hamiltonian also changes if one considers time-dependent gauge transformations, which is something to keep in mind when quantizing the theory. We will not pursue this further here, and we shall assume that $H$ is already determined. For a thorough discussion on the quantization of gauge theories, we refer the reader to Ref.~\cite{Henneaux:1992ig}.}, revealing the presence of a color octet $Q\bar{Q}$ pair. Outside this time interval the QGP evolves in the absence of external color sources.

Using Eq.~\eqref{eq:W-as-eiHt}, one can write:
\begin{align}
    & \frac{3 N_c}{g^2 T_F } [g_{\rm adj}^{++}]^>(t)  \\ 
    &= \frac{{\rm Tr}_{\mathcal{H}} \! \left[ e^{ i H t} E^a_i(0) \! \left[e^{- i (H - g A_0^c(0) T_{\rm adj}^c ) t}\right]^{ab} \!  E^b_i(0) e^{-\beta H} \right]}{{\rm Tr}_{\mathcal{H}} \left[  e^{-\beta H} \right]} \, , \nn
\end{align}
where the trace ${\rm Tr}_{\mathcal{H}}$ runs over physical states of the QGP. The analytic continuation is now direct, because all of the time dependence is in the exponentials. We just set $t \to - i\tau$, and identify the Euclidean gauge field $A_4$ with the Minkowski one by $A_0(0) = i A_4(0)$ (which in turn means that the electric field picks up a factor of $i$), to find
\begin{align}
    & [g_{\rm adj}^{++}]^>(-i\tau) \nn \\ 
    &= - \frac{g^2 T_F}{3 N_c}  \frac{{\rm Tr}_{\mathcal{H}} \! \left[ e^{ H \tau} E^a_i(0) \! \left[e^{-  (H - g A_0^c(0) T_{\rm adj}^c ) \tau }\right]^{ab} \! E^b_i(0) e^{-\beta H} \right]}{ {\rm Tr}_{\mathcal{H}} \left[  e^{-\beta H} \right] } \, \nn \\
    &= - \frac{g^2 T_F}{3 N_c}  \big\langle E_i^a(\tau) \! \left[ {\rm P} \exp \left( ig \int_0^\tau \diff \tau' \, A_4^c(\tau') T_{\rm adj}^c \right) \right]^{ab} \! E_i^b(0) \big\rangle_T \nonumber \\
    & = - \frac{g^2 T_F }{3 N_c} \big \langle E_i^a(\tau) W^{ab}(\tau,0) E_i^b(0) \big\rangle_T \nn\\
    &\equiv  G_{\rm adj}(\tau) \, , \label{eq:G-analytic-cont}
\end{align}
where ${\rm P}$ denotes path-ordering.
That is to say, we have proven that one of the real-time correlations we want to evaluate is related to an Euclidean correlation function by $ [g_{\rm adj}^{++}]^{>}(-i\tau) = G_{\rm adj}(\tau)$. We note that the absence of the denominator term as in Eq.~\eqref{eqn:Gfund} is a result of the absence of a Wilson line along the imaginary time direction at $t=-\infty$ in the definition of $[g_{\rm adj}^{++}]^{>}$. In quarkonium dissociation, the initial state is a color singlet, whereas in heavy quark diffusion, the initial state is in a color triplet representation, whose effect appears explicitly in the initial thermal state.

\subsection{Extraction of Quarkonium Transport Coefficients from Euclidean QCD}
Now we discuss how to extract the quarkonium transport coefficients from $G_{\rm adj}(\tau)$. Even though this correlation function has been studied in the past~\cite{Eidemuller:1997bb,Eidemuller:1999mx,DElia:2002hkf}, its precise connection to quarkonium transport has remained unexplored, until now. 
It turns out that neither Eq.~\eqref{eqn:fund_from_rho} nor Eq.~\eqref{eqn:Gfund_rho} is valid for the quarkonium case. This is so because Eq.~\eqref{eqn:fund_from_rho} is a result of the standard KMS relation, which, as we will show momentarily, is more complicated for the quarkonium correlator. Furthermore, Eq.~\eqref{eqn:Gfund_rho} relies on the spectral function being odd in $\omega$, which is crucially not true for the quarkonium correlator, as we will discuss in what follows. 

\subsubsection{KMS Relation and Non-odd Spectral Function}
\label{sect:non-odd}
To explain the non-oddness of the spectral function for quarkonium transport, we follow Ref.~\cite{Binder:2021otw} to use the usual proof of the KMS relation, plus the time-reversal operation and find
\be \label{eqn:standard_kms}
[g_{\rm adj}^{++}]^>(\omega) = e^{\omega/T} [g_{\rm adj}^{--}]^>(-\omega) \,,
\ee
which is the necessary KMS relation for proper thermalization of the internal degrees of freedom of the heavy quark pair (their relative motion and internal quantum numbers~\cite{Yao:2017fuc}). We then introduce the spectral function that governs quarkonium transport as
\be \label{eqn:rho_adj}
\rho_{\rm adj}^{++}(\omega) = [g_{\rm adj}^{++}]^>(\omega) - [g_{\rm adj}^{--}]^>(-\omega) \, ,
\ee
which, by definition satisfies $[g_{\rm adj}^{++}]^>(\omega) = (1 + n_B(\omega)) \rho_{\rm adj}^{++}(\omega)$, with $n_B(\omega) = (e^{\beta \omega}-1)^{-1}$. We have kept the superscripts ``$++$'' in the label of this spectral function because we can  also define
\be
\rho_{\rm adj}^{--}(\omega) = [g_{\rm adj}^{--}]^>(\omega) - [g_{\rm adj}^{++}]^>(-\omega) \, ,
\ee
which contains the same information, and satisfies $\rho_{\rm adj}^{--}(\omega) = - \rho_{\rm adj}^{++}(-\omega)$.

Here comes the most important part: The spectral function~\eqref{eqn:rho_adj} is not odd in $\omega$. In the standard thermal field theory setup, we define $\rho(\omega) = g^>(\omega) - g^<(\omega)$ where
$g^>(t) = \langle \phi(t) \phi(0) \rangle$ and
$g^<(t) = \langle \phi(0) \phi(t)\rangle$, which are related via $g^>(\omega) = g^<(-\omega)$ in frequency space by time translational invariance. This immediately leads to $\rho(\omega) = -\rho(-\omega)$. However, the relation $g^>(\omega) = g^<(-\omega)$ is not true for $[g_{\rm adj}^{++}]^>(\omega)$ and $[g_{\rm adj}^{--}]^>(\omega)$ due to the path ordering of field operators and the additional Wilson line along the imaginary time in $[g_{\rm adj}^{--}]^>$. That is to say, $[g_{\rm adj}^{--}]^>(t) \neq [g_{\rm adj}^{++}]^>(t)$. Therefore, we do not know how $\rho_{\rm adj}^{++}(\omega)$ transforms under $\omega\to-\omega$ a priori.

To see this more formally, one may also write the spectral function as a spectral decomposition in terms of the eigenvalues/eigenstates of $H$, denoted by $\{E_n, |n\rangle\}$, and those of $[H \mathbbm{1} - g A_0^c(0) T^c_{\rm adj} ]^{ab}$, denoted by $\{\tilde{E}_n , |\tilde{n}^a \rangle \}$, where $a$ is interpreted as a component of the state, rather than a label. With these definitions, it follows that
\begin{align}
    \rho_{\rm adj}^{++}(\omega) = \frac{g^2 T_F}{3 N_c} \sum_{n, \tilde{n}}  (2\pi) & \delta(\omega + E_n - \tilde{E}_{\tilde{n}})  | \langle n | E_i^a(0) | \tilde{n}^a \rangle |^2 \nn \\ & \quad   \times \left[ e^{-\beta E_n} - e^{-\beta \tilde{E}_{\tilde{n}} } \right] \, .
\end{align}
There is no reason why this expression would be odd under $\omega \to -\omega$, because the energies $E_n$ and $\tilde{E}_n$ can (and will) be different in general.

Indeed, explicit perturbative calculations at NLO show that $\rho_{\rm adj}^{++}(\omega)$ contains both $\omega$-odd, which is the usual case, and $\omega$-even parts (see Appendix~\ref{app:spectral}). The final result Eq.~(3.66) shown in Ref.~\cite{Binder:2021otw} is only for $\omega>0$, as mentioned there. We have performed a similar calculation for $\omega<0$ and found an $\omega$-even part, which originates from the diagrams (5, 5r) of Ref.~\cite{Binder:2021otw}, or diagrams (j) of Refs.~\cite{Eller:2019spw,Burnier:2010rp}
\begin{align} \label{eq:rho-diff}
\Delta \rho(\omega) \equiv \big(\rho^{++}_{\rm adj}(\omega) - \rho_{\rm fund}(\omega) \big) = \frac{g^4 T_F (N_c^2-1) \pi^2}{3 (2\pi)^3}  |\omega|^3 \,,
\end{align}
where we have also added a factor of $2$ since the definition of the spectral function shown in Eq.~(3.66) of Ref.~\cite{Binder:2021otw} differs from Eq.~\eqref{eqn:rho_adj} by a factor of $2$ (see Eq.~(3.28) therein). 

To demonstrate the importance of the $\omega$-even part, we use it to recompute the difference between $\gamma_{\rm fund}$ and $\gamma_{\rm adj}$ at the order of $\alpha_s^2$
\be
\label{eqn:delta_gamma}
\Delta \gamma \equiv \gamma_{\rm adj} - \gamma_{\rm fund} = - \frac{16 \zeta(3)}3 T_F C_F N_c  \alpha_s^2 T^3   \,,
\ee
where $C_F = \frac{N_c^2 - 1}{2N_c}$. This difference was first calculated in Ref.~\cite{Eller:2019spw}.  
Some algebra and use of the definitions for $[g_{\rm adj}^{\pm \pm}]^{>}$ leads to
\begin{align}
\gamma_{\rm adj} &= {\rm Im} \int_{-\infty}^{+\infty} \diff t \big( \theta(t) [g^{++}_{\rm adj}]^>(t) + \theta(-t) [g^{++}_{\rm adj}]^>(-t) \big) \nn\\
\Delta \gamma &= -\frac1{\pi}  \int_{-\infty}^{+\infty} \! \frac{\diff \omega }{|\omega|} \big(\theta(\omega) + n_B(|\omega|) \big) \Delta \rho(\omega) \,, 
\end{align}
where we have used $[g_{\rm adj}^{\pm \pm}]^{>}(\omega) = (1+n_B(\omega))\rho_{\rm adj}^{\pm \pm}(\omega)$ and used that they are translationally invariant in time.
The piece proportional to $\theta(\omega)$ is a pure vacuum contribution that vanishes in dimensional regularization. The second term inside the integral, however, gives a thermal contribution:
\begin{align} \label{eq:Delta-gamma-proof}
    \Delta \gamma &= - \frac{4 g^4 T_F}{3 (2\pi)^4} \pi^2 (N_c^2 - 1)  \int_0^{+\infty} \frac{\omega^2 \diff \omega}{e^{\omega/T} - 1} \\
    &= -\frac{16 \zeta(3)}{3} T_F C_F N_c \alpha_s^2  T^3 \, , \nn
\end{align}
which is exactly the difference given in Eq.~\eqref{eqn:delta_gamma}. This settles a long-standing issue regarding the consistency of the gauge-invariant chromoelectric correlators in the adjoint and fundamental representation, and verifies explicitly that the spectral function relevant for quarkonium transport is qualitatively different from that for heavy quark diffusion. The above discrepancy $\Delta \gamma$ is explained precisely because $\rho^{++}_{\rm adj}(\omega)$ is not odd in frequency.

With these theoretical foundations in hand, we can now proceed to write down the formula analogous to Eq.~\eqref{eqn:Gfund_rho}, which will allow for the extraction of $\kappa_{\rm adj}$ and $\gamma_{\rm adj}$ from the evaluation of the Euclidean correlator $G_{\rm adj}(\tau)$.

\subsubsection{Extraction Formulas}
Using the fact that $G_{\rm adj}(\tau)$ is the analytic continuation of $[g_{\rm adj}^{++}]^>(t)$ to Euclidean signature, we can write
\begin{align}
    G_{\rm adj}(\tau) &= \int_{-\infty}^{+\infty} \frac{\diff \omega}{2\pi} e^{-\omega \tau} [g_{\rm adj}^{++}]^{>}(\omega) \label{eq:G-adj-inversion} \\
    &=  \int_{-\infty}^{+\infty} \frac{\diff \omega}{2\pi} \frac{\exp \big( \omega ( \frac{1}{2T} - \tau ) \big)}{2\sinh \big( \frac{\omega}{2T} \big) } \rho_{\rm adj}^{++}(\omega) \, .\nn
\end{align}
However, in contrast to Eq.~\eqref{eqn:Gfund_rho}, the integrand may not be symmetrized with respect to $\omega$ because $\rho_{\rm adj}^{++}(\omega)$ is neither even nor odd. We note that, as one might suspect from Eq.~\eqref{eq:G-analytic-cont} and is apparent from Eq.~\eqref{eq:G-adj-inversion}, the analytic continuation holds provided that $0 < \tau < \beta$. This is precisely the range where we discuss the calculation of $G_{\rm adj}$ in the next section.
A direct calculation using Eqs.~\eqref{eqn:Gfund_rho},~\eqref{eq:rho-diff}, and~\eqref{eq:G-adj-inversion} shows that
\begin{align}
    \Delta G(\tau) &\equiv G_{\rm adj}(\tau) - G_{\rm fund}(\tau) \\ 
    &=  \int_{-\infty}^{+\infty} \frac{\diff \omega}{2\pi} \frac{\exp \big( \omega ( \frac{1}{2T} - \tau ) \big)}{2 \sinh \big( \frac{\omega}{2T} \big) } \Delta \rho (\omega)  \nonumber \\
    &= \frac{g^4 T_F (N_c^2 - 1)}{(2\pi)^3} \pi T^4 \big[ \zeta ( 4, \tau T ) - \zeta ( 4, 1 - \tau T ) \big] \nonumber \\ & \quad + \mathcal{O}(g^6) \, , \nonumber
\end{align}
where $\zeta(s,a) = \sum_{k=0}^\infty (k+ a)^{-s}$ is the Hurwitz zeta function.

After extracting $\rho_{\rm adj}^{++}(\omega)$ from the lattice QCD calculated $G_{\rm adj}(\tau)$, which will be discussed in the next section, we can obtain $\kappa_{\rm adj}$ and $\gamma_{\rm adj}$ as
\begin{align}
\label{eqn:extraction}
\kappa_{\rm adj} &= \lim_{\omega\to0} \frac{T}{2\omega} \left [\rho_{\rm adj}^{++}(\omega) - \rho_{\rm adj}^{++}(-\omega) \right] \\
\gamma_{\rm adj} &= - \int_0^\beta \diff \tau \, G_{\rm adj}(\tau)  \nn \\
& \quad - \frac{1}{2\pi} \int_{-\infty}^{+\infty} \!\!\! \diff \omega  \frac{1 + 2n_B(|\omega|)}{|\omega|} \rho_{\rm adj}^{++}(\omega) \,, \nn
\end{align}
where the expression we have written for $\kappa_{\rm adj}$ makes it manifest that only the $\omega$-odd part of $\rho_{\rm adj}^{++}(\omega)$ contributes to it. (One can show this by using Eqs.~\eqref{eqn:adj} and~\eqref{eqn:g>_to_gT}.)
We note that $\gamma_{\rm adj}$ may be substantially more difficult to extract than in the fundamental representation case. While the first term is indeed the same as in the fundamental case by virtue of $ \int_{-\infty}^{+\infty} \frac{\diff \omega}{2\pi} \frac{\rho_{\rm adj}^{++}(\omega)}{\omega} = \int_0^\beta \diff \tau G_{\rm adj}(\tau)$, the fact that $\rho_{\rm adj}^{++}$ is not necessarily odd under $\omega \to - \omega$ means that the last term can contribute. Indeed, it does so in perturbation theory, as demonstrated by our calculation of $\Delta \gamma$ in Eq.~\eqref{eq:Delta-gamma-proof}. There is even an additional complication in that the $1$ in $1+2n_B$ of the second line will usually generate ultraviolet divergences that have to be regulated analytically (e.g., by dimensional regularization).
Furthermore, the first term may also require regularization for the integration regions where $\tau \approx 0, \beta$.

\begin{figure*}[t]
    \centering
    \includegraphics[width=0.89\textwidth]{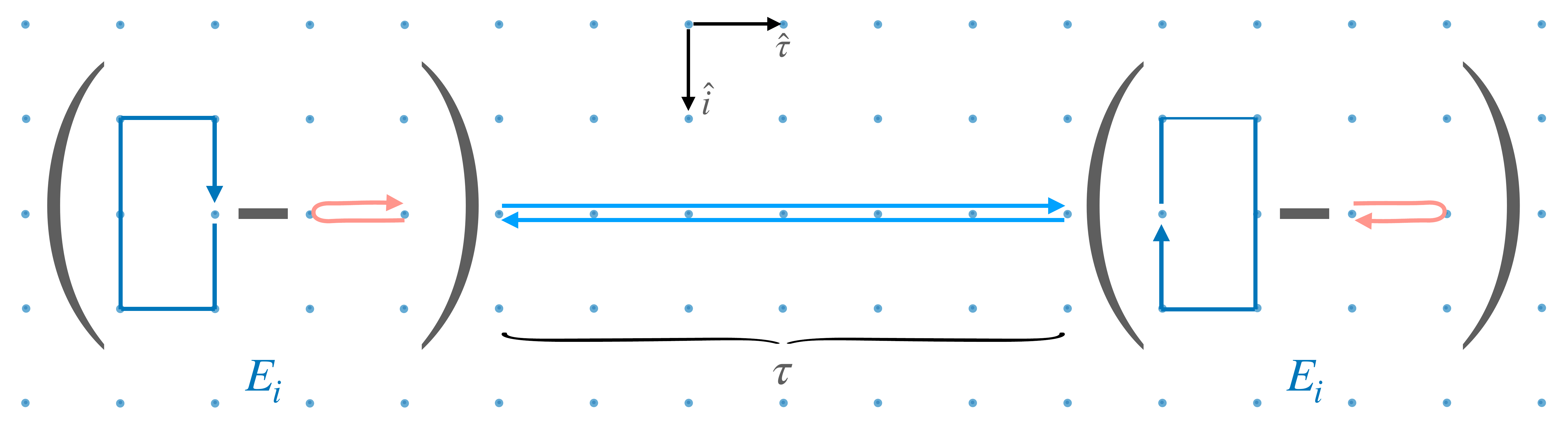}
    \caption{Lattice discretization of the chromoelectric field correlator. The electric field insertions are constructed by taking the difference between the products of gauge links over the blue and red contours at the ends of the light blue contours, which represents an adjoint Wilson line. In this setup, the adjoint Wilson line is equivalent to two antiparallel fundamental Wilson lines.}
    \label{fig:EE-picture}
\end{figure*}

\section{Lattice QCD Determination of $G_{\rm adj}(\tau)$ and Renormalization}
\label{sect:lattice}
In this section, we discuss how to perform a lattice QCD calculation of $G_{\rm adj}$ and extract $\rho_{\rm adj}^{++}$. We will first show a discretized version of $G_{\rm adj}$ and then discuss how to renormalize the lattice QCD result when taking the continuum limit. Finally we will give a fitting ansatz to extract $\rho_{\rm adj}^{++}$ from the calculated $G_{\rm adj}$, which can then be plugged into Eq.~\eqref{eqn:extraction} to obtain the quarkonium transport coefficients.

\subsection{Lattice Discretization}
The main ingredient we require in order to construct a lattice formulation of the correlator that determines quarkonium transition rates is a discretized formula for the gauge field strength $F_{\mu \nu} = \partial_\mu A_v - \partial_\nu A_\mu - i g [A_\mu, A_\nu ] $ in terms of link variables $U_\mu(n) = \exp (i a g A_\mu(n) )$ :
\begin{align}
    [\Delta U]_{\mu \nu}(n) &= U_{-\nu}(n + \hat{\nu} ) U_{-\mu}(n + \hat{\mu}+\hat{\nu}) U_\nu(n+\hat{\mu}) \nonumber \\ & \quad \times U_{\nu}(n+\hat{\mu}-\hat{\nu}) U_\mu(n - \hat{\nu}) U_{-\nu}(n) - 1 \nonumber \\ &= 2 i g a^2 F_{\mu \nu}(n) + \mathcal{O}(a^3) \, .
\end{align}
This discretization is different from the standard square plaquette. We chose this one because it makes the operator symmetric around the Wilson line direction, as shown in Fig.~\ref{fig:EE-picture}.
One can then write an expression purely in terms of link variables for the correlator:
\begin{align}
    G_{\rm adj}(\tau; a) &= \frac{(-1)}{12 a^4 N_c} \left\langle \! {\rm Tr}_{c} \! \left\{ \! \left( \prod_{n= n_\tau-1}^{0} U^\dagger_{0}(n) \right) \!  [\Delta U]_{\tau i}( n_\tau ) \right. \right. \nonumber \\ &  \quad  \left. \left. \times \! \left( \prod_{n=0}^{n_\tau - 1} U_{0}(n) \right) \!  [\Delta U]_{(-\tau) (-i)}( 0 ) \! \right\} \! \right\rangle_E   ,
\end{align}
where $\tau = a n_\tau$, and the products are ordered in such a way that the lower limit of the index labels corresponds to the operator that is most to the right in the product, and the upper limit to the one that is most to the left. A graphic representation of the correlator can be found in Fig.~\ref{fig:EE-picture}. The average $\langle \cdot \rangle_E$ represents the expectation value under the measure defined by the Euclidean lattice path integral, i.e., $\langle O \rangle_E = \frac{1}{\ml{Z}_E} \int DU \exp(- S_E[U]) O[U] $ where $\ml{Z}_E = \int DU \exp(- S_E[U])$.

\subsection{Renormalization and Infrared Renormalon}
The bare chromoelectric correlator $G_{\rm adj}(\tau; a)$ can be evaluated by the lattice method explained above. For physical quantities, the lattice calculation result needs proper renormalization. Since the operator involves a Wilson line, it is expected that $G_{\rm adj}(\tau; a)$ contains a linear divergence (which has not been explicitly checked and should be done so in the future via, e.g., a calculation in lattice perturbation theory), in addition to the usual logarithmic divergence. Therefore, we renormalize the bare correlator via
\begin{align}
\label{eqn:GR_adj}
G^R_{\rm adj}(\tau,\mu) = Z(\mu,a) e^{\delta m\cdot\tau} G_{\rm adj}(\tau; a) \,,
\end{align}
where $Z$ stands for the renormalization factor for the logarithmic divergence of the composite operator, with $\mu$ the renormalization scale and $\delta m$ the mass renormalization associated with the self energy of the Wilson line.

It has been shown that this form of the renormalization factor for the nonlocal operator is consistent with the fact that when the nonlocal operator is expressed as a weighted sum of local lattice operators, they mix in the renormalization group flow~\cite{Musch:2010ka}. In this work, we will not address the potential mixing between similar correlators with different Wilson line paths connecting the two chromoelectric fields.

A NLO calculation of the real-time partner of $G_{\rm adj}$, i.e., $[g_{\rm adj}^{++}]^>$ has shown that~\cite{Binder:2021otw}
\begin{align}
Z' = 1 + \frac{0}{\epsilon} + {\rm finite\ terms\ at\ }g^2 +\ml{O}(g^4) \,,
\end{align}
where we used $Z'$ to distinguish the renormalization factor for $[g_{\rm adj}^{++}]^>$ from the $Z$ for $G_{\rm adj}$. The ``$0$'' coefficient of the $1/\epsilon$ term emphasizes that $[g_{\rm adj}^{++}]^>$ has no logarithmic divergence at NLO.
The calculation was performed in the continuum by using dimensional regularization. The divergent term should be the same in the dimensionally regularized and lattice regularized perturbative calculations. Only the finite terms can be different. If we want to obtain the renormalized result in the $\overline{\rm MS}$ scheme, the finite difference between the lattice scheme result and the $\overline{\rm MS}$ result should still be accounted for. In the case of $G_{\rm fund}$, the difference is known at NLO~\cite{Christensen:2016wdo}. We leave the calculations of $Z$ for the Euclidean $G_{\rm adj}$ in both schemes to future studies. (As can be seen by comparing to Ref.~\cite{Christensen:2016wdo}, such calculations are research projects on their own.)

Since the $\delta m$ term is associated with the self energy of the Wilson line, one can use lattice perturbative calculations to determine it, but the uncertainties are expected to be large due to infrared renormalons. In particular, $\delta m$ is expected to be of the form
\begin{align}
\delta m = \frac{m_{-1}(a \Lambda_{\rm QCD})}{a} + m_0(\Lambda_{\rm QCD})\,,
\end{align}
where $m_{-1}$ is constant at leading order in lattice perturbation theory, but it has a residual dependence on $a$ at higher orders via, e.g., $a\Lambda_{\rm QCD}$ due to renormalization effects. On the other hand, $m_0$ is independent of the lattice spacing $a$, but it is scheme dependent as well. (Both $m_{-1}$ and $m_0$ also depend on the other mass scales of the theory, if there are any.) The infrared renormalon ambiguity leads to an uncertainty in summing the perturbative series for $m_{-1}$, which is compensated by the same uncertainty in determining $m_0$. The fact that both $m_{-1}$ and $m_0$ are scheme dependent is reflected in the systematic uncertainty of fitting the $a$ dependence from lattice calculations at small $a$, as shown in the recent study on renormalizing the quasi parton distribution function (quasi-PDF)~\cite{LatticePartonCollaborationLPC:2021xdx}.

Here we discuss a strategy to reduce the uncertainty caused by the infrared renormalons in determining the renormalization factor $\delta m$ by using lattice QCD calculation results, which is motivated by the recent work on self renormalization of the quark quasi-PDF~\cite{LatticePartonCollaborationLPC:2021xdx,Zhang:2023bxs}. The first step is to fit $m_{-1}$ from the $a$ dependence of $Z(\mu, a)G_{\rm adj}(\tau;a)$ when $a$ is small for some $\tau$. Different choices of $\tau$ are expected to give the same fitting result, as long as we maintain $\tau\gg a$ to have negligible lattice artifacts). Due to the unknown nonperturbative dependence of $m_{-1}$ on $a$, different parametrizations may be used in the fitting and they do not lead to the same result necessarily, which reflects the scheme dependence of $m_{-1}$. Then we define $G^{R'}_{\rm adj}(\tau,\mu) \equiv Z(\mu, a) e^{m_{-1} \tau/a} G_{\rm adj}(\tau;a)$, i.e., we only absorb the extracted $a$-dependent linear divergence and the logarithmic divergence into the renormalization factor and perform an operator production expansion (OPE) at small $\tau$ (i.e., $\beta \gg \tau$ but we still require $\tau \gg a$)
\begin{align}
\label{eqn:remove_IRR}
& G^{R'}_{\rm adj}(\tau,\mu) = e^{-m_0\tau} \sum_n C_n(\alpha_s(\mu), \mu\tau) \tau^n \langle O_n \rangle_T^R(\mu)  \\
& \xrightarrow{\tau\to 0} (1-m_0\tau) \sum_{n=0,1} C_n \tau^n \langle O_n \rangle_T^R  + \ml{O}(\tau^2) \,, \nn 
\end{align}
where $O_n$ denotes the local operators in the OPE and $\langle O_n \rangle_T^R(\mu)$ represents their renormalized expectation values at the same temperature $T$. The expectation values of $O_n$ can be calculated by standard lattice QCD methods and renormalized perturbatively by calculating the corresponding logarithmic renormalization factors via lattice perturbative calculations, in the same way as it is done for the logarithmic renormalization factor $Z$ for $G_{\rm adj}$. These local operators do not involve Wilson lines and thus do not have linear divergence, so it is expected that their renormalization is insensitive to the effects from infrared renormalons. The local OPE operators that may contribute include
\begin{align}
O_0&:\quad \mathbb{1}\,,\ \Tr_{\rm c}(F_{0i}F_{0i})\,,\ \Tr_{\rm c}(F_{ij}F_{ij})\,,\ m_q\bar{q}q \\
O_1&:\quad e_\rho \Tr_{\rm c}(F_{0i} D^\rho F_{0i})\,,\ e_\rho \Tr_{\rm c}(F_{ij} D^\rho F_{ij})\,,\ e_\rho m_q\bar{q}D^\rho q \,, \nn
\end{align}
where $e_\rho$ is a unit vector along the spacetime direction $\rho$. The short-distance Wilson coefficients $C_n$ can be calculated in perturbation theory at the scale $\mu=1/\tau$. The calculation of these coefficients is an active area of research~\cite{Braun:2020ymy,Braun:2021cqe}. In practice, we can determine $m_0$ via Eq.~\eqref{eqn:remove_IRR} by calculating the lattice renormalized $G^{R'}_{\rm adj}(\tau,\mu)$ and $\langle O_n \rangle_T^R(\mu)$. With $m_0$ determined, we can obtain $G_{\rm adj}^R(\tau,\mu)$ from $G^{R'}_{\rm adj}(\tau,\mu)$ by including the renormalization factor associated with $m_0$. As suggested in Ref.~\cite{Zhang:2023bxs}, to reduce the uncertainty caused by the infrared renormalons, one resums the leading infrared renormalons in $C_n$ by regulating the renormalon poles in the Borel space and applying the inverse Borel transformation. As shown therein, this strategy removes a large uncertainty in the determination of the quark PDF. We expect a similar uncertainty reduction to happen for the determination of $G_{\rm adj}^R$ by using this strategy.

After determining the renormalized $G_{\rm adj}^R$ in the lattice regularization, we can convert it into the $\overline{\rm MS}$ scheme if we know the difference between the perturbative results of the logarithmic divergence in these two schemes. As part of the conversion process, one has to take care of the fact that in dimensional regularization with $d=4-\epsilon$ and $\epsilon\to0$, the linear divergence is absent. Any residual finite terms from this linear divergence are accounted for through $m_0$ in the OPE matching.

\subsection{Fitting Ansatz for $\rho_{\rm adj}^{++}$}
Once we obtain the renormalized $G_{\rm adj}^R$, we can use Eq.~\eqref{eq:G-adj-inversion} to fit the spectral function $\rho_{\rm adj}^{++}$. Since we only have a limited number of data points in $\tau$, we need a fitting ansatz. One ansatz that has been used in the lattice studies of the heavy quark diffusion coefficient is of the form~\cite{Altenkort:2023oms}
\begin{align}
\label{eqn:ansatz}
\rho_{\rm adj}^{++}(\omega) = \sqrt{\rho_{\rm IR}^2(\omega) + \rho_{\rm UV}^2(\omega)} \,,
\end{align}
where $\rho_{\rm IR}$ and $\rho_{\rm UV}$ represent ansatzes in the small and large $\omega$ regions, respectively. We will construct ansatzes motivated from perturbative studies.

The $\omega$-even part of the large frequency behavior of $\rho_{\rm adj}^{++}(\omega)$ is determined by Eq.~\eqref{eq:rho-diff}. The remaining ($\omega$-odd) terms can be read off directly from~\cite{Binder:2021otw}, where $\rho_{\rm adj}^{++}(\omega)$ was calculated at $\omega > 0$.
Explicitly,
\begin{align}
\label{eqn:rho_UV}
    \rho^{++}_{\rm adj}(\omega)  \overset{\omega \gg T}{=} \, &  \frac{g^2 T_F (N_c^2-1) \omega^3 }{ 3  \pi N_c } \times   \\  \bigg\{ 1 & + \frac{g^2}{(2\pi)^2} \bigg[ \left( \frac{11 N_c}{12} - \frac{N_f}{6} \right) \ln \left( \frac{\mu^2}{4 \omega^2} \right) \nonumber \\ &  + N_c \left( \frac{149}{36} - \frac{\pi^2}{6} + \frac{\pi^2}{2} {\rm sgn}(\omega) \right) - \frac{5 N_f}{9}  \bigg] \bigg\} \nonumber \\ &+ \mathcal{O}(g^6) \, , \nonumber
\end{align}
where $N_f$ is the number of light (massless) quark flavors in the theory. It was shown in Ref.~\cite{Burnier:2010rp} that up to $\mathcal{O}(g^4)$, the leading temperature-dependent contributions (which are the same for $\rho_{\rm adj}^{++}$ and $\rho_{\rm fund}$, cf.~\cite{Binder:2021otw}) at large frequency go as $T^4/\omega$, which are omitted in Eq.~\eqref{eqn:rho_UV} since they are subleading.

On the infrared side, one needs to use the hard thermal loop effective theory to capture the behavior of correlation functions when $|\omega| \lesssim g T \propto m_D$, where $m_D$ is the so-called Debye mass of the QGP, given (perturbatively) by $m_D^2 = g^2 T^2 \left( \frac{N_c}{3} + \frac{N_f}{6} \right) $, which quantifies color-electric screening in a thermal plasma. To see the difference between the $\rho_{\rm fund}$ and $\rho_{\rm adj}$ in the small $\omega$ region, one needs to consider the same type of diagrams that led to the difference shown in Eq.~\eqref{eq:rho-diff}, which has a prefactor of $g^4$, meaning that the dominant corrections in the regime $|\omega| \lesssim m_D$ will be of order $g^4 m_D^2 |\omega| \propto g^6 T^2 |\omega|$. This means that we cannot make quantitative statements by considering only the 1-loop diagram that leads to Eq.~\eqref{eq:rho-diff} (replacing the propagators with their HTL-resummed counterparts), as we can get competing effects from 2-loop diagrams in QCD, which contribute at order $g^6$. In practice, one would also need to calculate these 2-loop diagrams to be able to match the HTL result to full QCD. We will leave such calculations to future studies. Here we only list the leading contribution in the infrared regime, which can be written in terms of the well-known heavy quark diffusion coefficient $\kappa_{\rm fund}$ at NLO:
\begin{align}
    \rho^{++}_{\rm adj}(\omega)  \overset{\omega \ll gT}{=} \rho_{\rm fund}(\omega)  \overset{\omega \ll gT}{=} \frac{\kappa_{\rm fund} \omega}{ T} + \mathcal{O}(g^6) \, , 
\end{align}
where $\kappa_{\rm fund}$ is given by~\cite{Caron-Huot:2007rwy,Caron-Huot:2008dyw}:
\begin{align}
    \kappa_{\rm fund} &= \frac{g^4 T_F (N_c^2-1) T^3 }{9 (2\pi) N_c} \times \\ & \quad \bigg[ \left( N_c + \frac{N_f}{2} \right) \left( \ln \frac{2 T}{m_D}  + \frac12 - \gamma_E + \frac{\zeta'(2)}{\zeta(2)} \right) \nonumber \\ & \quad \quad + \frac{N_f}{2} \ln 2 + \frac{N_c m_D}{T} C \bigg] + \mathcal{O}(g^6) \nonumber \, ,
\end{align}
with $C \approx 2.3302$, as given in Ref.~\cite{Caron-Huot:2008dyw}. The fact that the low-frequency limit of the adjoint and fundamental correlators do not differ up to this order had already been noticed in Ref.~\cite{Caron-Huot:2008dyw}.

Motivated by the above perturbative analyzes, we suggest to use Eq.~\eqref{eqn:rho_UV} as $\rho_{\rm UV}$ in the fitting ansatz~\eqref{eqn:ansatz} and use $\kappa_{\rm adj} \omega + c |\omega|$ to parametrize $\rho_{\rm IR}$ with $c$ some constant that does not contribute to $\kappa_{\rm adj}$. The appearance of the $c |\omega|$ term in $\rho_{\rm IR}$ is a crucial difference from the case of the heavy quark diffusion coefficient and is motivated by perturbative calculations shown in Section~\ref{sect:non-odd}. The fitting of $\rho_{\rm adj}^{++}$ will not only provide the quarkonium transport coefficient $\kappa_{\rm adj}$, but also the frequency dependence of $\rho_{\rm adj}^{++}$, which is important to evaluate $\gamma_{\rm adj}$, as well as the frequency-dependent correlators $g^{\pm \pm}_{\rm adj}(\omega)$ that determine the quarkonium dissociation and recombination rates.

\section{Conclusions}
\label{sect:conclusion}
In this paper, we explained how to determine the real time quarkonium transport properties from a Euclidean chromoelectric field correlator. This determination requires to reconstruct a spectral function in a way that is different from more intensively studied spectral function reconstruction problems, such as the one required for the extraction of the heavy quark diffusion coefficient. The key results are shown in Eq.~\eqref{eqn:extraction}. We then discussed the lattice determination of the Euclidean correlator, and in particular, a method to reduce the uncertainty caused by infrared renormalons in obtaining the renormalization factor for the linear divergence of the correlator. This method is quite involved and several perturbative calculations needed to implement the method are left to future studies, such as the lattice-regularized perturbative calculation of the logarithmic renormalization factor $Z$ in Eq.~\eqref{eqn:GR_adj} and the Borel-resummed calculation of the Wilson coefficients in the OPE~\eqref{eqn:remove_IRR}. Our work paves a way towards a nonperturbative determination of the quarkonium transport properties in the QCD hot medium, which generalizes the use of a weakly interacting gas of quarks and gluons as a microscopic model of the QGP in Boltzmann (rate) equations~\cite{Rapp:2017chc,Yao:2018sgn,Du:2019tjf,Yao:2020xzw} for quarkonium to the strongly coupled case.
This not only deepens our understanding of the QGP and quarkonium production in heavy ion collisions, but may also provide insights for studies of exotic heavy flavor production~\cite{Yao:2018zze,Wu:2022blx,Wu:2023djn} and dark matter bound state formation in the early universe~\cite{Binder:2020efn,Binder:2021otw,Biondini:2023yxt,Biondini:2023zcz}.

\vspace{0.5cm}

After publication of this work, we noticed that although the difference $\Delta \rho$ between spectral functions we found in this work in Eq.~\eqref{eq:rho-diff} reproduces the difference between the transport coefficients $\gamma_{\rm adj}$ and $\gamma_{\rm fund}$ present in~\cite{Eller:2019spw}, it does not reproduce the difference between the coefficients of the $\pi^2$ terms in the result of the calculation for $\rho_{\rm adj}^{++}$ in~\cite{Binder:2021otw} and the result of the Euclidean QCD calculation of $\rho_{\rm fund}$ in~\cite{Burnier:2010rp}. In the original version of this work, we made a transcription error that led us to find no tension. It is clear from earlier calculations by others~\cite{Eidemuller:1997bb} and ourselves~\cite{Binder:2021otw} that the $\omega > 0$ part of the spectral function $\rho_{\rm adj}^{++}$ at $T=0$ is as given in Eq.~\eqref{eqn:rho_UV}. It is also clear that the only $\omega$-even contributions to $\rho_{\rm adj}^{++}$ are given by Eq.~\eqref{eq:rho-diff}, as $\rho_{\rm fund}$ is odd in $\omega$ and Eq.~\eqref{eq:rho-diff} unambiguously represents the subtraction of the $\omega$-odd contribution from diagrams (5) and (5r) of~\cite{Binder:2021otw}. Therefore, the $\omega < 0$ part of $\rho_{\rm adj}^{++}$ at $T=0$ can be cross-checked with the above two calculations and is as given in Eq.~\eqref{eqn:rho_UV}. There is no thermal contribution to the difference at 1-loop because this difference can be written in terms of the imaginary part of free retarded correlators, which do not depend on $T$. To unambiguously establish the origin of the discrepancy between the $\pi^2$ terms in $\rho_{\rm adj}^{++}$ and $\rho_{\rm fund}$, and as an additional cross-check, we have calculated the difference between $G_{\rm adj}$ and $G_{\rm fund}$ in the imaginary time formalism and added it as Appendix~\ref{app:euclidean} to this work. We find that the difference between $\rho_{\rm adj}^{++}$ and $\rho_{\rm fund}$ is as given in Eq.~\eqref{eq:rho-diff}, and therefore, that the term proportional to $\pi^2$ in $\rho_{\rm fund}$ should be $-\pi^2/6$ instead of $-2\pi^2/3$. That is to say, we find that in~\cite{Burnier:2010rp}, $-8\pi^2/3$ in Eq.~(4.2) should be $-2\pi^2/3$. This new calculation will appear as part of the erratum in the journal version.

\begin{acknowledgments}
We thank Xiangdong Ji, Joshua Lin, Yin Lin, Guy D. Moore, Patrick Oare, Peter Petreczky, Krishna Rajagopal, Martin J. Savage, Phiala E. Shanahan and Iain W. Stewart for useful comments. We thank Mikko Laine for invaluable discussions that helped us determine the origin of the discrepancy between the results for $\rho_{\rm fund}$ and $\rho_{\rm adj}^{++}$ present in the literature.
We thank the Institute for Nuclear Theory (INT) at the University of Washington for its kind hospitality and stimulating research environment, and the organizers of the INT-22-3 program ``Heavy Flavor Production in Heavy‐Ion and Elementary Collisions''. This research was supported in part by the INT's U.S. Department of Energy grant No. DE-FG02-00ER41132.
B.S. is supported by the U.S. Department of Energy, Office of Science, Office of Nuclear Physics under grant Contract Number DE-SC0011090. X.Y. also acknowledges support from the U.S. Department of Energy, Office of Science, Office of Nuclear Physics, InQubator for Quantum Simulation (IQuS) under Award Number DOE (NP) Award DE-SC0020970.
\end{acknowledgments}

\bibliography{main.bib}

\begin{widetext}
\newpage
\appendix
\section{Detailed Expressions in the Lindblad Equation}
\label{app:lindblad}
Here we write out explicitly each term in the Lindblad equation~\eqref{eqn:lindblad} introduced in the main text, which can be found in the literature, e.g., in Ref.~\cite{Brambilla:2022ynh}.
The density matrix is assumed to be block diagonal in the color singlet and octet basis
\begin{align}
\rho_S(t) = \begin{pmatrix}
\rho_S^{(s)}(t) & 0 \\
0 & \rho_S^{(o)}(t)
\end{pmatrix} \,.
\end{align}
The Hamiltonian and its thermal correction are given by [$C_F=(N_c^2-1)/(2N_c)$]
\begin{align}
H_S = \frac{{\bs p}_{\text{rel}}^2}{M} + \begin{pmatrix} 
- \frac{C_F\alpha_s}{r} & 0 \\
 0 & \frac{\alpha_s}{2N_cr}
\end{pmatrix} \,, \qquad\quad \gamma_{\rm adj} \Delta h_S = \frac{\gamma_{\rm adj}}{2} r^2 \begin{pmatrix}
1 & 0\\
0 & \frac{N_c^2-2}{2(N_c^2-1)}
\end{pmatrix} \,,
\end{align}
The Lindblad operators are given by
\begin{align}
L_{1i} &= \Big(r_i + \frac{1}{2MT}\nabla_i - \frac{N_c}{8T}\frac{\alpha_s r_i}{r} \Big)\begin{pmatrix}
0 & 0\\
1 & 0
\end{pmatrix}  \\
L_{2i} &= \sqrt{\frac{1}{N_c^2-1}}\Big(r_i + \frac{1}{2MT}\nabla_i + \frac{N_c}{8T}\frac{\alpha_s r_i}{r} \Big)\begin{pmatrix}
0 & 1\\
0 & 0
\end{pmatrix} \nn \\
L_{3i} &= \sqrt{\frac{N_c^2-4}{2(N_c^2-1)}}
\Big(r_i + \frac{1}{2MT}\nabla_i \Big)\begin{pmatrix}
0 & 0\\
0 & 1
\end{pmatrix} \,, \nn
\end{align}
where $i=x,y,z$.

\section{Calculation Details of Spectral Function Difference}
\label{app:spectral}

As explained in the main text, the difference between the spectral function for quarkonium transport and that for single heavy quark transport is given by the diagrams (j) in Refs.~\cite{Burnier:2010rp,Eller:2019spw}, or (5), (5r) in Ref.~\cite{Binder:2021otw}. The diagrammatic representation of their difference in real time in terms of Wightman functions was given in Ref.~\cite{Scheihing-Hitschfeld:2022xqx}, where gauge invariance was also examined.

Following the calculation details of Ref.~\cite{Binder:2021otw}, we find that the difference between these two spectral functions stemming from these diagrams is given by
\begin{align}
    \rho^{++}_{\rm adj}(\omega) -  \rho_{\rm fund}(\omega) = \int_{{\bs p}, k} & \frac{T_F}{3N_c} g^4 N_c (N_c^2 - 1)  2\pi \delta(k_0)  \big[ g_{\mu \nu} (p - 2k)_\delta + g_{\nu \delta} (k - 2p)_{\mu} + g_{\delta \mu} (p+k)_\nu \big]  \\ 
    \times &  (p_0 g_{i\delta'} - p_i g_{0\delta'} ) \big((p_0-k_0) g_{i\nu'} - (p_i - k_i) g_{0\nu'} \big) \nonumber \\
    \times &  {\rm Re} \Big\{  [\rho(p)]^{\delta' \delta} [D_T(p-k)]^{\nu \nu'} [D_T(k)]^{\mu 0}  \nonumber \\
    & \quad \quad  - [D_{T}(p)]^{\delta' \delta } \big( [D_>(p-k)]^{\nu' \nu } [D_>(k)]^{ 0 \mu}  - [D_<(p-k)]^{\nu' \nu } [D_<(k)]^{\mu 0}  \big) \Big\} \, , \nn
\end{align}
where $p_0 = \omega$. By using the thermal (KMS) relations between the free propagators $D_>, D_<, D_T$ and $\rho$, this can be further simplified to
\begin{align}
\label{eq:general-spectral-diff-NLO}
    \rho^{++}_{\rm adj}(\omega) -  \rho_{\rm fund}(\omega) &= \int_{{\bs p}, k}  \frac{T_F}{3N_c} g^4 N_c (N_c^2 - 1)  2\pi \delta(k_0)  \big[ g_{\mu \nu} (p - 2k)_\delta + g_{\nu \delta} (k - 2p)_{\mu} + g_{\delta \mu} (p+k)_\nu \big]  \\ 
    & \quad \times (p_0 g_{i\delta'} - p_i g_{0\delta'} ) \big((p_0-k_0) g_{i\nu'} - (p_i - k_i) g_{0\nu'} \big) \nonumber \\
    & \quad \times (-1) [\rho(p)]^{\delta' \delta} {\rm Im} \{ [D_R]^{\nu \nu'}(p-k) \} {\rm Im} \{ [D_R]^{\mu 0}(k) \} \, .  \nn
\end{align}
In our convention, the free propagators in Feynman gauge are given by
\begin{align}
    [\rho(p)]^{\mu \nu} = (- g^{\mu \nu}) (2\pi) {\rm sgn}(p_0) \delta(p^2) & & [D_R(p)]^{\mu \nu} = \frac{ - i g^{\mu \nu}}{p^2 + i 0^+ {\rm sgn}(p_0) } \, ,
\end{align}
and using them to calculate the difference, one arrives at
\begin{align}
    \rho^{++}_{\rm adj}(\omega) -  \rho_{\rm fund}(\omega) &= \int_{{\bs p}, k}  \frac{T_F}{3N_c} g^4 N_c (N_c^2 - 1)  (2\pi) \delta(k_0)   (2\pi) {\rm sgn}(\omega) \delta(p^2) \mathcal{P} \left( \frac{2 d \omega^3 - 2 \omega ({\bs p} - {\bs k})^2}{k^2 (p-k)^2 } \right)  \, .
\end{align}

In dimensional regularization, $({\bs p} - {\bs k})^2$ may be exchanged by $\omega^2$ because $\int_{\bs k} \frac{1}{{\bs k}^2}$ vanishes. Then, setting $d = 3$, this integral becomes
\begin{align}
    \rho^{++}_{\rm adj}(\omega) -  \rho_{\rm fund}(\omega) =  \frac{T_F}{3N_c} g^4 N_c (N_c^2 - 1) |\omega|^3   \int_{{\bs p}, {\bs k}}   (2\pi) \delta(p^2) \mathcal{P} \left( \frac{(-4) }{{\bs k}^2 [ \omega^2 - ({\bs p} - {\bs k})^2 ] } \right)  \, .
\end{align}
The explicit calculation of this integral is equivalent to the one presented in the Supplemental Material of Ref.~\cite{Scheihing-Hitschfeld:2022xqx}. The final result is
\begin{align}
    \rho^{++}_{\rm adj}(\omega) -  \rho_{\rm fund}(\omega) &=  \frac{T_F}{3N_c} g^4 N_c (N_c^2 - 1) |\omega|^3   \frac{\pi^2}{(2\pi)^3}  = \frac{g^4 T_F (N_c^2 - 1) \pi^2}{3 (2\pi)^3} |\omega|^3 \, ,
\end{align}
as claimed in the main text.

It is noteworthy that the difference between the spectral functions, as given in Eq.~\eqref{eq:general-spectral-diff-NLO} may also be used in conjunction with HTL-resummed propagators to explore the value of the difference (a modification to the gluon 3-vertex is also necessary, according to the HTL effective theory Feynman rules. They can be found in Ref.~\cite{Andersen:2002ey}.). However, as discussed in the main text, a full fixed-order calculation at $\mathcal{O}(g^6)$, which is the leading contribution to the difference in the small frequency domain, also requires considering 2-loop diagrams, which we will not pursue here.

\section{Calculation Details of Euclidean Correlators Difference}
\label{app:euclidean}

In this Appendix we report the results of a direct calculation of the difference between the Euclidean correlators that characterize heavy quark diffusion~\eqref{eqn:Gfund} and quarkonium transport~\eqref{eq:G-analytic-cont}. A direct calculation in terms of Feynman diagrams shows that the difference up to $\mathcal{O}(g^4)$ is given by
\begin{equation}
    \Delta G(\tau) = G_{\rm adj}(\tau) - G_{\rm fund}(\tau) = - \frac{g^3 T_F}{6 N_c} f^{abc} \big\langle \mathcal{T}_E \big( \partial_4 A_i^a(\tau) - \partial_i A_4^a(\tau) \big) \int_0^\beta {\rm d}\tau' A_4^c(\tau') \big( \partial_4 A_i^b(0) - \partial_i A_4^b(0) \big) \big\rangle_{\mathcal{O}(g)} \, .
\end{equation}
The subscript $\mathcal{O}(g)$ indicates that only the tree-level 3-gluon vertex contributes, and $\mathcal{T}_E$ denotes Euclidean time ordering (terms with bigger imaginary time arguments are implicitly pushed to the left of the expression).
It is interesting to see that the Matsubara zero mode of the gauge field appears explicitly in these expressions.

A direct calculation in dimensional regularization (DR), introducing Feynman parameters when appropriate, leads to
\begin{align}
    \Delta \tilde{G}(k_n) = -i  \frac{g^4 C_F N_c}{3} k_n^3 \times \big(k_n^2\big)^{D-4}  \times \left\{ \frac{\Gamma \! \left(\frac{3-D}{2} \right)^2 }{2 (4\pi)^{D-1}} + (D - 2) \frac{\Gamma(4-D)}{(4\pi)^{D-1}} \int_0^1 \!\! {\rm d}x \int_0^1 \!\! {\rm d}y \, \frac{\left[ \frac{1-y+xy}{\sqrt{y(1-y+yx(1-x))}} \right]^{D-4}}{\sqrt{y} (1 - y + yx(1-x))^{3/2} }   \right\} \, ,
\end{align}
where all of the dependence on $k_n$ is outside the curly bracket, and the terms inside the curly bracket simply correspond to a numerical prefactor.

One can show that $\Delta \rho(\omega) = 2 {\rm Im} \left\{ \Delta \tilde{G} (k_n) \right\}_{k_n \to -i(\omega + i0^+)}$. Performing the analytic continuation $k_n \to -i(\omega + i0^+)$ means that we obtain
\begin{equation}
    -i k_n^3 \times \big(k_n^2\big)^{D-4} \to  \omega^3 \times \big( -\omega^2 - i \omega 0^+ \big)^{D-4} = \omega^3 |\omega|^{2D-8}  \times e^{-i \pi {\rm sgn}(\omega) (D-4)} \, ,
\end{equation}
where the analytic continuation from $k_n$ to $\omega$ is taken by continuously deforming $k_n$ starting from the real axis into the imaginary axis, without actually crossing the imaginary axis (i.e., without crossing the negative $k_n^2$ axis).
Its imaginary part is
\begin{equation}
    {\rm Im} \{  \omega^3 |\omega|^{2D-8} \times e^{-i \pi {\rm sgn}(\omega) (D-4)} \} = \pi (4-D) {\rm sgn}(\omega) \omega^3 |\omega|^{2D-8} + \mathcal{O}((D-4)^3) \, .
\end{equation}

It then follows that the difference between spectral functions in the limit $D \to 4$ is purely determined by the divergent contribution to $\Delta \tilde{G}(k_n)$. In the limit, $(4 - D) \Gamma(4-D) \to 1$, and we may set $D = 4$ elsewhere. The integral over Feynman parameters gives a simple result
\begin{equation}
    \int_0^1 \!\! {\rm d}x \int_0^1 \!\! {\rm d}y \, \frac{1}{\sqrt{y} (1 - y + yx(1-x))^{3/2} } = 2\pi \, ,
\end{equation}
with which
\begin{align}
    \Delta \rho(\omega) = \lim_{D \to 4} 2 \, {\rm Im} \left\{ \Delta\tilde{G}\big(-i (\omega + i0^+) \big) \right\}
    = \frac{g^2 C_F \omega^3}{3\pi} \frac{g^2}{(2\pi)^2} N_c \frac{\pi^2}{2} {\rm sgn}(\omega) \, ,
\end{align}
just as we obtained via our real time calculation.

\renewcommand{\thesubsection}{C.\arabic{subsection}}
\subsection{Resolving the tension with previous results}

We now discuss the calculation in~\cite{Burnier:2010rp} of the terms proportional to $\pi^2$, and how they arrived at a different result. In short, the issue is that the IR regulators employed in~\cite{Burnier:2010rp} fundamentally alter the analytic structure of the integral to be calculated.

In~\cite{Burnier:2010rp}, the integral structure that generates the $\pi^2$ terms is given by their Eq.~(A.42)
\begin{equation}
    \delta_{3m} \tilde{\mathcal{I}}_5 = \frac{32 \pi^3}{(4\pi)^2}  \int_{\bs k} {\rm Im} \left\{ \int_0^{1/2} \!\!\! {\rm d}s \frac{k_n^2}{k_n^2 + k^2 } \frac{2k_n^2 }{(2 s k_n )^2 + k^2} \left( \ln \frac{k_n^2 + k^2}{k_n^2} - \ln (1-4s^2) \right) \right\}_{k_n \to -i\omega + 0^+} \, , \label{eq:delta3m}
\end{equation}
where we have omitted the DR scale $\mu$, and written the expression without the IR regulator $\lambda$ present in their work (they write $\frac{k_n^2}{k_n^2 + k^2 + \lambda^2}$ instead of $\frac{k_n^2}{k_n^2 + k^2}$ next to the $\diff s$ integral sign). We have also written $2k_n^2$ in the place of $k_n^2 - k^2$ (the numerator on the second fraction under the $s$ integral sign) because one can show that their difference will not lead to any terms proportional to $\pi^2$ in the result. Furthermore, we have multiplied their expression by $16 \pi^3$ so that it contributes to $\rho_{\rm fund}$ as the numerical factor obtained from~\eqref{eq:delta3m} that multiplies $2 g^4 T_F N_c (N_c^2-1) \omega^3/(3(2\pi)^3 )$. That is to say, the result of the multiplication of~\eqref{eq:delta3m} with $2 g^4 T_F N_c (N_c^2-1) \omega^3/(3(2\pi)^3 )$ is an additive contribution to $\rho_{\rm fund}$. (Because of all of these changes we denote the first symbol as $\delta_{3m}$ instead of $\delta_3$.)

\begin{figure}
    \centering
    \includegraphics[width=\textwidth]{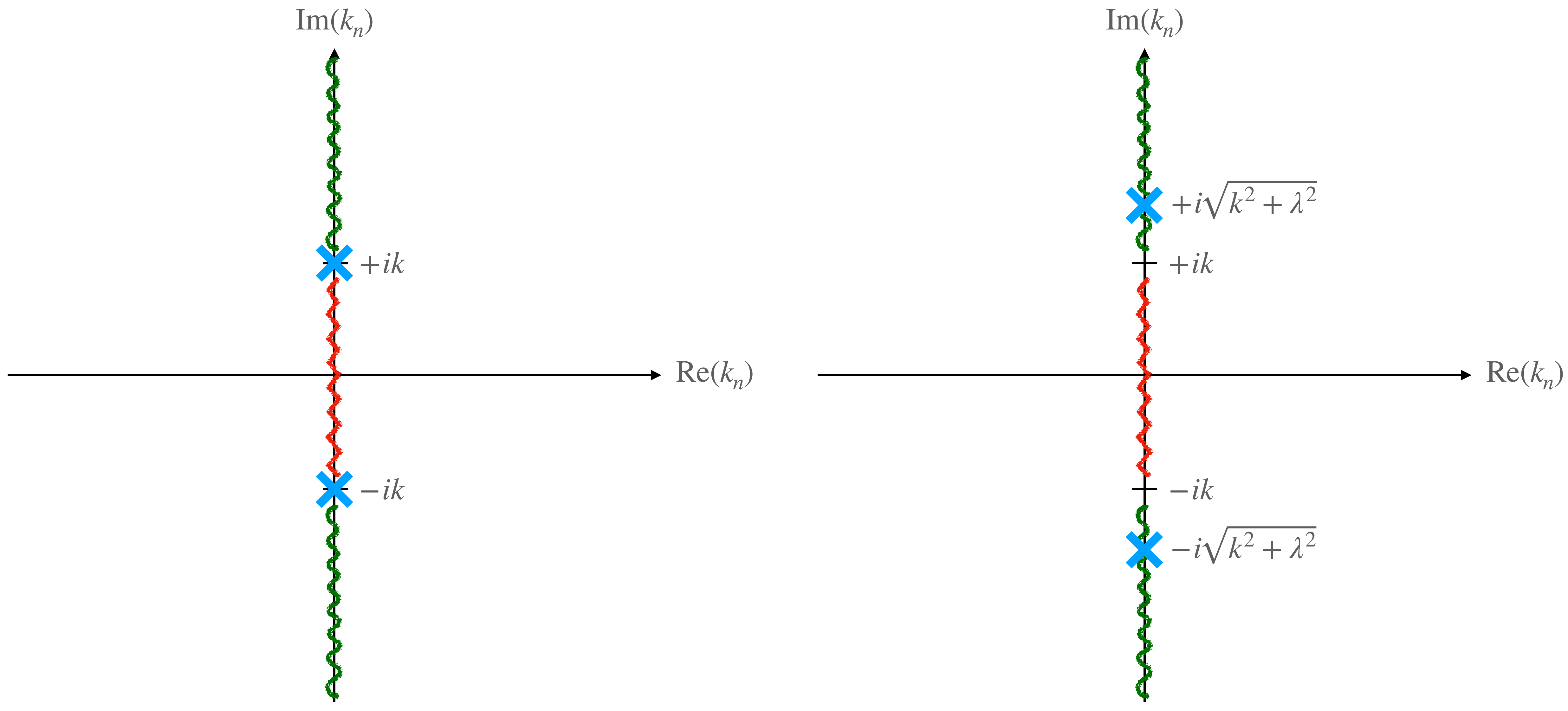}
    \caption{Graphic representation of the pole structure of Eq.~\eqref{eq:delta3m} in the complex $k_n$ plane at fixed $k$. Poles are represented with blue crosses, the branch cut between $-ik$ and $+ik$ is represented by a red zig-zag line, and the branch cuts above and below $\pm ik$ are represented by wavy green lines. The latter branch cut is induced by the presence of the Wilson line. Left: The pole structure without a regulator. Right: The pole structure with the regulator used in~\cite{Burnier:2010rp}. Without the branch cuts induced by the Wilson line, this regulator is not problematic because the branch cut denoted by a red zig-zag line does not intersect the poles. However, with the branch cuts induced by the Wilson line, moving the poles in this manner qualitatively alters the pole structure, because the contributions from the regions where $ i k < \pm {\rm Im}(k_n) < i \sqrt{k^2 + \lambda^2} $, ${\rm Re}(k_n) \approx 0$ will contribute with an opposite sign to the unregulated version.}
    \label{fig:poles}
\end{figure}

The calculation of~\cite{Burnier:2010rp} proceeds by introducing a regulator in the form of a mass term, then doing the analytic continuation, taking the imaginary part, and evaluating the integrals at the end. This would work if such a regulator did not change the positions of the poles relative to the branch cuts of the integrand, which, crucially, it does. If we view the integrand of Eq.~\eqref{eq:delta3m} as a function of $k_n$ in the complex plane,
at each fixed $k$, there are poles at $k_n=\pm i k$, branch cuts starting at $k_n = \pm i k$ and extending to $\pm i\infty$ due to the integration over $s$, and a branch cut between $k_n = \pm i k$ due to the explicit logarithm in the integrand. See Fig.~\ref{fig:poles} for a graphic representation.

Starting from this picture, introducing a regulator in the denominator of the first factor under the $s$ integral sign amounts to moving the positions of the poles into the branch cut generated by the integration over $s$. Since the analytic continuation is essentially a limit from the right in Fig.~\ref{fig:poles}, it is crucial that the position of the poles relative to the branch cuts be faithful to the observable one intends to calculate. The regulator in~\cite{Burnier:2010rp} does not satisfy this requirement. Indeed, one can verify by a direct numerical calculation that
\begin{equation}
    \delta_{3m} \tilde{\mathcal{I}}_5(\omega) = \frac{\pi^2}{3} {\rm Im} \left\{  k_n^2 \sqrt{k_n^2} \right\}_{k_n \to -i\omega + 0^+} \, ,
\end{equation}
as opposed to $(-2\pi^2/3) {\rm Im} \{  k_n^2 \sqrt{k_n^2} \}_{k_n \to -i\omega + 0^+}$, which is what was found in~\cite{Burnier:2010rp}.

Furthermore, there is an additional contribution that the calculation in Appendix A.4 of~\cite{Burnier:2010rp} did not consider, which to our knowledge was first calculated in~\cite{Eller:2021qpp} (see pages 154-160). It originates explicitly from the Matsubara zero mode.
This contribution that was neglected in~\cite{Burnier:2010rp} corresponds to $-\pi^2/2$ in our normalization of the terms in the parenthesis with the prefactor $N_c$ in Eq.~\eqref{eqn:rho_UV}. It then follows that the term proportional to $\pi^2 $ in the sought result is $\pi^2/3 - \pi^2/2 = - \pi^2/6$, as we claimed earlier.

\end{widetext}

\end{document}